\documentclass[aps, reprint, showpacs, epsfig]{revtex4-1}

\usepackage{array,longtable}
\usepackage{graphics}
\usepackage{graphicx}
\usepackage{subfigure}
\usepackage{epsfig}
\usepackage{latexsym}
\usepackage{epstopdf} 
\usepackage{color}
\usepackage{amsmath}
\usepackage{mathtools}

\usepackage{hyphenat}
\let\oldref\ref
\renewcommand{\ref}[1]{(\oldref{#1})}
\usepackage[colorlinks=true,linkcolor=blue,urlcolor=blue,citecolor=blue]{hyperref}
\usepackage{amsmath}
\usepackage[normalem]{ulem}

\begin{document}
\title{Emergence of cooperatively reorganizing cluster and super-Arrhenius dynamics of fragile
supercooled liquids}

\author{Ankit Singh$^{1}$}
\author{Sarika Maitra Bhattacharyya$^{2}$}
\author{Yashwant Singh$^{1}$}
\affiliation{$^{1}$Department of Physics, Banaras Hindu University, Varanasi-221 005, India.}

\affiliation{$^{2}$Polymer Science and Engineering Division, CSIR-National Chemical Laboratory, Pune-411008, India}
\date{\today}

\begin{abstract}
In this paper we develop a theory to calculate the structural relaxation time $\tau_{\alpha}$
of fragile supercooled liquids. Using the information of the configurational entropy and 
structure we calculate the number of dynamically free, metastable, and stable neighbors around 
a central particle. In supercooled liquids the cooperatively reorganizing clusters (CRCs) in 
which the stable neighbors form ``stable'' nonchemical bonds with the central particle emerge.
For an
event of relaxation to take place these bonds have to reorganize irreversibly; the energy involved
in the processes is the effective activation energy of relaxation. The theory brings forth a 
temperature $T_{a}$ and a temperature dependent parameter $\psi(T)$ which characterize slowing
down of dynamics on cooling. It is shown that the value of $\psi(T)$ is equal to $1$ for $T>T_{a}$
indicating that the underlying microscopic mechanism of relaxation is dominated by the entropy 
driven processes while for $T<T_{a}$, $\psi(T)$ decreases on cooling indicating the 
emergence of the energy driven processes. 
This crossover of $\psi(T)$ from high to low
temperatures explains the crossover seen in $\tau_{\alpha}$. The dynamics of 
systems that may have similar static structure but very different dynamics can be understood 
in terms of $\psi(T)$. We present results for the Kob-Anderson model for three densities and 
show that the calculated values of $\tau_{\alpha}$ are in excellent agreement with simulation
values for all densities. We also show that when $\psi(T)$, $\tau_{\alpha}$ and other 
quantities are plotted as a function of $T/T_{a}$ (or $T_{a}/T$) the data collapse on master curves.
\end{abstract}


\maketitle

\section{Introduction\label{Intro}}
When a liquid is supercooled bypassing its crystallization, it continues to 
remain structurally disordered, but its dynamics slows down so quickly that 
below a temperature, called the glass temperature $T_{g}$,
the structural relaxation takes such a long time that it becomes almost 
impossible to observe \cite{Angell}. The structural relaxation time, $\tau_{\alpha}$,
represents the time required for the liquid to return to equilibrium after a small
perturbation. The super-Arrhenius temperature dependence of $\tau_{\alpha}$
(or the viscosity) is the defining characteristic of a `fragile' liquid \cite{Mallamace, Martinez}.
The super-Arrhenius behavior suggests that the effective activation energy for
the relaxation in a fragile liquid increases with decreasing temperature. 
The underlying reason for such a behavior is understood in terms of increasing 
the cooperativity of relaxation on cooling. The cooperativity can be defined in 
terms of number of particles which move in some sort of concert in order for
an elementary relaxation event to occur. These particles may be distributed 
in a region without forming a compact structure and share the space with other 
particles \cite{stevenson_Wolynes_nature,Rajesh_ganapathy_PRL}. 
In such a situation it would not be possible to characterize cooperativity 
in terms of a spatial length. Alternatively, one can think of a compact cooperative 
region defined by a length $\xi$; the number of particles in the region varies as $\xi^{d}$,
where d is spatial dimension \cite{Adam, Kirkpatrick}. To have a precise picture of cooperativity 
(or the cooperative region), its determination as a function of temperature and 
calculation of the effective activation energy from microscopic interactions 
between particles have been the major focus of the theoretical description of 
supercooled liquids and the glass transition \cite{Adam, Kirkpatrick, Ediger, Kivelson, Garrahan, LG, Paddy_JCP_143_044503_2015, Coslovich, Tanaka, Mosayebi, Smarajit_Indrajit_Tah, Bouchaud,  Biroli, Montanari, Hocky, Cammarota, Kurchan, Sausset}.

Is there any local structural order in a liquid that grows rapidly on cooling and can be 
associated with the cooperativity? This question has been the subject of much activity of
last several years \cite{Paddy_JCP_143_044503_2015, Coslovich, Tanaka, Mosayebi, Smarajit_Indrajit_Tah, Bouchaud,  Biroli,  Montanari, Hocky, Cammarota, Kurchan, Sausset} 
as its resolution would lead to a better understanding of glassy 
phenomena and would provide insights into the underlying microscopic mechanism of the
relaxation. However, liquid structure determined from scattering experiments that give 
information at the level of two-point correlation functions, such as the structure 
factor and the pair correlation function shows no such local order. 
This led to the conclusion that if at all there is any local order linked to the cooperativity,
it would have to be very subtle and hidden to the pair correlation function.
This spurred several proposals of local preferred structure such as the 
``point-to-set length'' \cite{Bouchaud, Biroli, Montanari, Hocky} and the ``patch correlation length'' \cite{Cammarota,Sausset, Kurchan} with
varying success. The main difficulty is that one does not know how to distinguish
an amorphous ordered structure from the one that exists in a normal liquid \cite{Ediger}.

In the Adam and Gibbs theory \cite{Adam} as well as in the random first order transition
(also known as the mosaic) theory \cite{Kirkpatrick, xia_Wolynes_PNAS, Lubchenko_wolynes_JCP} the cooperative region is expressed in terms 
of the configurational entropy. Adam and Gibbs visualized a supercooled liquid as 
progressively organizing in larger and larger cooperative regions that have to 
collectively reorganize and proposed that the number of particles in a ``cooperatively
rearranging region (CRR)'' is inversely proportional to the configurational entropy $S_{c}$.
Since the configurational entropy $S_{c}(T)$ decreases on lowering the temperature, the 
number of particles in the cooperative region increases. The relaxation time $\tau_{\alpha}$
is the time needed to rearrange the region and is given by 

\begin{subequations}
\renewcommand{\theequation}{\theparentequation.\arabic{equation}}
\begin{align}
\tau_{\alpha}=\tau_{0} \mathrm{exp}{\left[\frac{A}{TS_{c}(T)}\right]}  ,
\end{align}
where A is a temperature independent phenomenological parameter, and $\tau_{0}$ is the 
microscopic timescale.

On the other hand, the mosaic theory assumes nucleation of so-called ``entropic droplets''
between different metastable configurations that makes the supercooled liquid as a 
patchwork of local metastable configurations. A static length $\xi$ in terms of which the 
size of the droplet is expressed is shown to be $\xi=\left(Y(T)/TS_{c}\right)^{\frac{1}{d-\theta}}$
where $d$ is the spatial dimension, $\theta$ is an exponent related to the interface energy
and $Y(T)$ is the surface tension \cite{xia_Wolynes_PNAS,Lubchenko_wolynes_JCP}. The relaxation time is related to the 
configurational entropy by relation,

\begin{equation}
\tau_{\alpha}=\tau_{0} \mathrm{exp} \left[\frac{1}{T}\lbrace\frac{Y(T)}{TS_{c}(T)}\rbrace^{\phi}\right]  ,
\end{equation}
where $\phi=\frac{d}{d-\theta}$. In order to find values of $\xi(T)$ and $\tau_{\alpha}(T)$
one has to know values of $Y(T)$ and $\theta$.

It has recently been shown \cite{Singh} that in a supercooled (supercompressed) liquid some 
particles get localized in potential wells and form long-lived (stable) nonchemical bonds
between them. A cluster of these bonded particles collectively rearranges and creates an 
effective potential energy barrier for relaxation. The number of particles in the cluster 
is calculated from the data of pair correlation function. In this paper, we extend the theory 
and report results for a glass-forming liquid.

The system we consider is the
Kob-Anderson 80:20 mixture of Lennard-Jones particles consisting of two species of particles a and b
\cite{Kob}. All particles have the same mass m and the interaction between two particles of type
$\alpha$,$\gamma$ $\in$ $[a,b]$ is given by
\begin{equation}
 u_{\alpha\gamma}(r)= 4\epsilon_{\alpha\gamma} 
[(\frac{\sigma_{\alpha\gamma}}{r})^{12}-(\frac{\sigma_{\alpha\gamma}}{r})^{6}] ,
\label{potential}
\end{equation}
\end{subequations}
with $\epsilon_{aa}=1$, $\sigma_{aa}=1$, $\epsilon_{ab}=1.5$, $\sigma_{ab}=0.8$, $\epsilon_{bb}=0.5$,
$\sigma_{bb}=0.88$. Length, energy and temperature are given in units of $\sigma_{aa}$, $\epsilon_{aa}$
and $\epsilon_{aa}/k_{B}$, respectively. The particles momenta of both species have identical
Maxwell-Boltzmann distribution.

\section{Number of ``stable'' and ``metastable'' bonds formed by a particle with its neighbors in a liquid \label{Theory}}
In equilibrium, each particle of a liquid feels on the average similar potential energy
barrier due to its interactions with neighbors, but its kinetic energy has a probability to
have any value given by the Maxwell-Boltzmann distribution. Therefore; due to competition
between the kinetic energy which makes particles to move and  the potential energy barrier 
that restricts particle motion, particles in a liquid acquire a wide range of dynamical
states. It is intuitively clear that all those particles whose values of kinetic energy are
high would be able to overcome the potential barrier and move around as free particles 
whereas all those particles whose values of kinetic energy are low would be trapped 
by the potential barrier. There are also particles whose values of kinetic energy are not
high enough, but fluctuations embedded in the system may make them escape 
the barrier. We can therefore roughly divide particles into three groups of dynamical states;
$(i)$ free particles who move around and collide with other particles,
$(ii)$ particles who remain trapped (localized) and execute vibrational motions at well
defined locations and $(iii)$ particles who are intermittent between trapped and free.
The concentration of these particles depends on density and temperature, the potential
energy barrier becomes higher on increasing the density and lowering the temperature 
and the kinetic energy of particles decreases on decreasing the temperature.
A supercooled liquid can therefore be considered as a network of particles connected 
with each other by (nonchemical) bonds with life-time varying from microscopic to
macroscopic time. We now describe how to calculate the number of these particles from 
the data of radial distribution function.

\subsection{Separation of $g(r)$ into parts representing particles of different dynamical states}

The radial distribution function which for a simple liquid is defined as \cite{Hansen}
\begin{subequations}
\renewcommand{\theequation}{\theparentequation.\arabic{equation}}
\begin{equation}
g(\lvert\vec{r_{2}}-\vec{r_{1}}\rvert) \equiv g(r) = 
\frac{1}{N\rho} \langle \sum_{j}^{N}\sum_{j \neq k}^{N} 
\delta(\vec{r}-\vec{r_{j}}+\vec{r_{k}}) \rangle ,  
\label{gr}
\end{equation}
where $N$ is number of particles, $\rho$, the number density and the angular bracket denotes the 
ensemble average, tells us what is probability of finding a particle at a distance $r$ from a 
reference (central) particle. The average number of particles lying within the range $r$ and
$r+\mathrm{d}r$ from the central particle in $3$-dimensions is $4\pi\rho g(r)r^{2}\mathrm{d}r$.
Since $g(r)$ defined by Eq.~(\ref{gr}) has no information about particle momenta, one can not say
how many of these particles located in the region at a given time will remain there forever
unless disturbed and how many of them will subsequently move away. To get such information
we define $g(r)$ of a binary mixture in the center-of-mass coordinates as \cite{Singh}, 

\begin{equation}
 g_{\alpha\gamma}(r)=\left(\frac{\beta}{2\pi\mu}\right)^{\frac{3}{2}}\int \mathrm{d}{\bf p} \  
\mathrm{e}^{-\beta (\frac{p^2}{2\mu} + w_{\alpha\gamma}(r))} , 
\end{equation}
where $\beta=(k_{B}T)^{-1}$ is the inverse temperature measured in units of 
the Boltzmann constant $k_{B}$, ${\bf p}$  is the relative momentum of a particle of mass 
$\mu=m/2$. The effective potential (potential of meanforce) $w_{\alpha\gamma}(r)=-k_{B}T\ln g_{\alpha\gamma}(r)$ 
is sum of the (bare) pair potential energy $u_{\alpha\gamma}$ and the system-induced potential 
energy of interaction between a pair of particles of species $\alpha$ and $\gamma$
separated by distance $r$ \cite{Hansen}.
The peaks and troughs of $g_{\alpha\gamma}(r)$ create, respectively, minima and maxima in $\beta w_{\alpha\gamma}(r)$
as shown in Fig. 1 for species $a$ at $T=0.45$ and $\rho=1.20$.
A region between two maxima, leveled as $i-1$ and $i$ $(i\geq 1)$ is denoted as $i$th shell
and minimum of the shell as $\beta w_{\alpha\gamma}^{(id)}$.
The value of $i$th maximum is denoted as $\beta w_{\alpha\gamma}^{(iu)}$ and its location by $r_{ih}$.

\begin{figure}[t]
\includegraphics[scale=0.4]{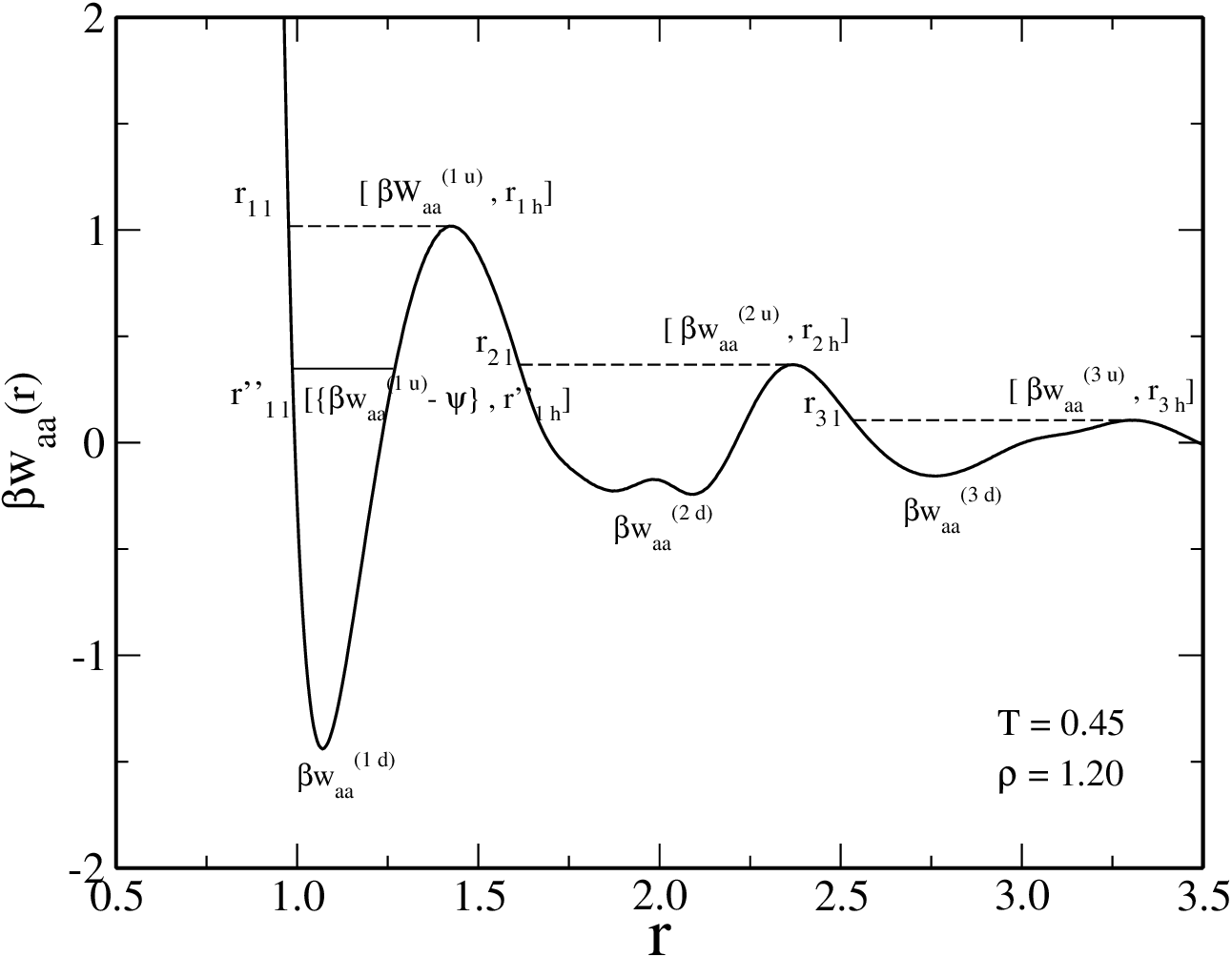}
\caption{The reduced effective potential $\beta w_{aa}(r)$ between a pair of particles of species $\alpha$
and $\gamma$ separated
by distance r (expressed in unit of $\sigma_{aa}$) in a system of Lennard-Jones at a
density $\rho=1.20$ and temperature $T=0.45$. $\beta w_{aa}^{(iu)}, r_{ih}$ are, respectively,
value and location of $i$th maximum and $r_{il}$ is the location on the left hand side of the shell where
$\beta w_{aa}^{(i)}(r)=\beta w_{aa}^{(iu)}$ (shown by dashed line).
The location $r''_{il}$ and $r''_{ih}$ are values of r on the left and the right hand side of the shell
where $\beta w_{aa}^{(i)}(r)=[\beta w_{aa}^{(iu)}-\psi]$ (shown by full line). $\beta w_{aa}^{(id)}$ is
the depth of the $i$th shell.}

\label{fig-1}
\end{figure}

In a classical system all those particles in region of $i$th shell whose energies are 
less or equal to $\beta w_{\alpha\gamma}^{(iu)}$
\textit{i.e.}, $\beta[\frac{p^2}{2\mu}+w_{\alpha\gamma}^{(i)}(r)] \leq \beta w_{\alpha\gamma}^{(iu)}$ 
will be trapped as they do not have enough energy to overcome the barrier $\beta w_{\alpha\gamma}^{(iu)}$.
These particles can be considered to be bonded with the central particle.
On the other hand, all those particles whose energies are higher than 
$\beta w_{\alpha\gamma}^{(iu)}$ or momenta higher than ${\sqrt{2\mu w_{\alpha\gamma}^{(iu)}}}$
are free to move around individually and collide with other particles as long as their momenta
remain higher than the value mentioned above. When due to collisions a free particle loses its momentum and 
falls below ${\sqrt{2\mu w_{\alpha\gamma}^{(iu)}}}$, the particle gets trapped.
At a given temperature and density an equilibrium between free and bonded particles is established.
The average number of these particles can be found from $g(r)$.

The averaged number of particles that form bonds with the central particle can be found from a part of
$g_{\alpha\gamma}(r)$ defined as

\begin{eqnarray}\nonumber
g_{\alpha\gamma}^{(ib)}(r) &=& 4\pi(\frac{\beta}{2\pi\mu})^{3/2} \mathrm{e}^{-\beta w_{\alpha\gamma}^{(i)}(r)} 
\int_{0}^{\sqrt{2\mu[w_{\alpha\gamma}^{(iu)}-w_{\alpha\gamma}^{(i)}(r)]}}      \\
&&  \ \times \mathrm{e}^{-\beta p^2/2\mu} p^2 \mathrm{d}p ,
\end{eqnarray}
where $w_{\alpha\gamma}^{(i)}(r)$ is the effective potential in the range of $r_{il}\leq r \leq r_{ih}$
of $i$th shell. Here $r_{il}$ is value of $r$ where $w_{\alpha\gamma}^{(i)}(r)= w_{\alpha\gamma}^{(iu)}$
on the left hand side of the shell (see Fig.~\ref{fig-1}). The number of particles in the shell which 
form  bonds with the central particle of species $\alpha$ is
\begin{equation}
n_{\alpha}^{(b)} = 4\pi \sum_{i} \sum_{\gamma}\rho_{\gamma}\int_{r_{il}}^{r_{ih}} 
g_{\alpha\gamma}^{(ib)}(r) r^2 \mathrm{d}r  ,
\end{equation}
where summations are over all shells and over all species and $\rho_{\gamma}$ is number density
of $\gamma$ species. This number $n_{\alpha}^{(b)}$ increases rapidly on lowering the temperature and increasing 
the density due to an increase in the number of shells surrounding the central particle
and increase in values of maximum and minimum of each shell. 

Since these bonded particles in each shell have a wide range of energies lying between
maximum and minimum of the shell, they oscillate with a wide range of frequencies.
However, all those particles whose energies are close to the maximum (barrier height)
may not remain bonded for long due to fluctuations embedded in the system (bath) 
which drive them to escape the barrier. This is an entropy driven process.
In case of an athermal system where only packing constraints matter, there are no
energy parameters whatsoever, the bath drives all those particles of $i$th shell
whose energies lie between $\beta w_{\alpha\gamma}^{(i u)}-1$ and  $\beta w_{\alpha\gamma}^{(i u)}$ out of the shell \cite{Singh}.
But in a thermal system, the entropy driven process is opposed by the energy
driven process; the system gains entropy but  loses internal energy when particles escape the shell
and reverse happens when particles remain in the shell. This competition results in a diminishing bath role in driving particles out of shells. 
This led us to introduce a temperature dependent parameter $\psi(T)$ $(\leq1)$
such that only those particles of $i$th shell whose energy lie between 
$\beta w_{\alpha\gamma}^{(i u)}-\psi$ and  $\beta w_{\alpha\gamma}^{(i u)}$ are able to escape the shell.
However, as is well known, the value of $\psi$ in a normal (high temperature)
liquid is one. The departure from $1$ is excepted to take place at lower 
temperatures where the role of energy parameters become important.
In Sec.~\ref{Theory}$B$ we describe a method to determine its value.

The bonded particles can be divided into two groups, one that consists of particles who 
are able to escape from the shell and the ones which survive fluctuations and remain
bonded unless disturbed. The first group of particles is all those particles whose energies lie between
$[\beta w_{\alpha\gamma}^{(iu)} - \psi]$ and $\beta w_{\alpha\gamma}^{(iu)}$ and momenta between
${\sqrt{2\mu[w_{\alpha\gamma}^{(iu)}-\psi k_{B}T- w_{\alpha\gamma}^{(i)}(r)]}}$ and
${\sqrt{2\mu[w_{\alpha\gamma}^{(iu)}-w_{\alpha\gamma}^{(i)}(r)]}}$. These particles are called metastable particles
and henceforth referred to as $m$-particles. During the time they remain bonded they oscillate
in the shell with time periods depending upon their energies. It is obvious that trajectories
of these particles are composed of a succession of periods of time when particles simply vibrate
around well defined locations (shells), separated by widely distributed in time, rapid jumps.
This feature has been observed in the computer simulation study of time-resolved square 
displacements of individual particles. The  plateau observed at intermediate times in the
mean-squared displacement is due to vibrations within shells.

The part of $g(r)$ that represents m-particles in the $i$th shell can be written as 
\begin{eqnarray}\nonumber
g_{\alpha\gamma}^{(im)}(r) &=& 4\pi(\frac{\beta}{2\pi\mu})^{3/2} \mathrm{e}^{-\beta w_{\alpha\gamma}^{(i)}(r)} 
\int_{\sqrt{2\mu[w_{\alpha\gamma}^{(iu)}-\psi k_{B} T- w_{\alpha\gamma}^{(i)}(r)]}}^{\sqrt{2\mu[w_{\alpha\gamma}^{(iu)}-w_{\alpha\gamma}^{(i)}(r)]}}      \\
&&  \ \times \mathrm{e}^{-\beta p^2/2\mu} p^2 \mathrm{d}p ,
\end{eqnarray}

Value of $g_{\alpha\gamma}^{(im)}(r)$ in a shell (see Fig. 5) starts from zero at 
$r=r'_{il}$ on the left hand side and attains a maximum value at a value of $r$ where 
$\beta w_{\alpha\gamma}^{(i)}(r)$ has its minimum value and then decreases and becomes zero
at $r'_{ih}$ on the right hand side of the shell. The number of m particles around a particle
of $\alpha$ species is found from $g_{\alpha\gamma}^{(im)}(r)$ using the relation,

\begin{equation}
n_{\alpha}^{(m)} = 4\pi \sum_{i} \sum_{\gamma}\rho_{\gamma}\int_{r'_{il}}^{r'_{ih}} 
g_{\alpha\gamma}^{(im)}(r) r^2 \mathrm{d}r  ,
\end{equation}

A length scale which can be associated with a cluster of m-particles formed around a central
particle is equal to the range of $g_{\alpha\gamma}^{(m)}(r) \simeq g_{\alpha\gamma}(r)-1$.
This length increases on lowering the temperature and increasing the density.
The averaged number of m-particles surrounding a central particle in a binary mixture is

\begin{equation}
n^{(m)} = x_{a} n_{a}^{(m)} + x_{b} n_{b}^{(m)}  ,
\end{equation}
where $x_{\alpha}$ is the concentration of species $\alpha$.

The second group of bonded particles are those whose energies are lower than 
$[\beta w_{\alpha\gamma}^{(iu)} - \psi]$ and particles momenta lower than
${\sqrt{2\mu[w_{\alpha\gamma}^{(iu)}-\psi k_{B} T- w_{\alpha\gamma}^{(i)}(r)]}}$. 
These particles form stable bonds with the central particle and are referred to as 
$s$-particles. The part of $g(r)$ that corresponds to these particles is 
\begin{eqnarray}\nonumber
g_{\alpha\gamma}^{(is)}(r) &=& 4\pi(\frac{\beta}{2\pi\mu})^{3/2} \mathrm{e}^{-\beta w_{\alpha\gamma}^{(i)}(r)} 
\int_{0}^{\sqrt{2\mu[w_{\alpha\gamma}^{(iu)}-\psi k_{B} T-w_{\alpha\gamma}^{(i)}(r)]}}      \\
&&  \ \times \mathrm{e}^{-\beta p^2/2\mu} p^2 \mathrm{d}p ,
\label{grs}
\end{eqnarray}
where $w_{\alpha\gamma}^{(i)}(r)$ is in the range $r''_{il}\leq r \leq r''_{ih}$. 
Here $r''_{il}$ and $r''_{ih}$ are, respectively, value of $r$ on the left and the 
right hand side of the shell where $\beta w_{\alpha\gamma}^{(i)}(r)= \beta w_{\alpha\gamma}^{(iu)}-\psi$. 
The number of particles around a $\alpha$ particle is 
\begin{equation}
n_{\alpha}^{(s)} = 4\pi \sum_{i} \sum_{\gamma}\rho_{\gamma}\int_{r''_{il}}^{r''_{ih}} 
g_{\alpha\gamma}^{(is)}(r) r^2 \mathrm{d}r  ,
\label{ns}
\end{equation}

The averaged number of s-particle bonded with a central particle in a binary mixture is

\begin{equation}
n^{(s)} = x_{a} n_{a}^{(s)} + x_{b} n_{b}^{(s)}  ,
\label{nst}
\end{equation}

We have to know value of $\psi$ as a function of $T$ and $\rho$ to calculate value of
$n^{(s)}(T)$ and $n^{(m)}(T)$ in a given system.

\subsection{Determination of temperature dependence of $\psi$ and the number of particles
in a cooperatively reorganizing cluster}

We call the cluster formed by $n^{(s)}$, $s$-particles bonded with the central particle 
as a cooperatively reorganizing cluster (CRC).
For an event of structural relaxation to take place, the cluster
has to reorganize irreversibly; the energy involved in this rearrangement is the
energy with which the central particle is bonded with $s$-particles. 
These particles are distributed in shells around the central particle where they share the
region with other (mobile) particles. As the structure of the cluster is not compact
it cannot be measured in terms of spatial length. The CRC, therefore, differs from
the Adam and Gibbs \cite{Adam} ``cooperatively rearranging region (CRR)''
which is taken to be a compact structure \cite{Adam, Bouchaud}.
The cooperativity here is defined in 
terms of number of bonds, $n^{(s)}$, formed by a particle with its neighbors.
As the temperature is lowered, the number $n^{(s)}$ as well as the energy of each bond in the cluster 
would increase. As a consequence, the relaxation time growth with decreasing temperature
is a super-Arrhenius \textit{i.e.}, faster than an exponential in inverse temperature.

Following Adam and Gibbs \cite{Adam} we assume that the relation between the  number of particles in a CRC 
and  the configurational entropy can be written as,
 
\begin{equation}
n^{(s)}(T)+1 = \dfrac{K}{S_{c}(T)} \ \ ,
\label{Sc}
\end{equation}
\end{subequations}
where K is a temperature independent constant and $S_{c}$ is the configurational entropy
per particle of the system. Values of $S_{c}$ as  function of temperature are found from 
relation $S_{c}(T)=S_{total}(T)-S_{vib}(T)$ where $S_{total}$ is sum of the ideal gas
entropy plus excess entropy arising due to interactions between particles and 
$S_{vib}$ is the vibrational entropy arising due to short-time vibrational motions
in a local potential energy minimum. Ingenious simulation techniques developed recently
\cite{Sastry, Berthier} have made  it possible to find accurate values of $S_{c}$ in supercooled region.
In present calculations we use values of $S_{c}$ reported in Ref. \cite{ABanerjee}.

\begin{figure*}[t]
\vspace{-1cm}
\includegraphics[scale=0.5]{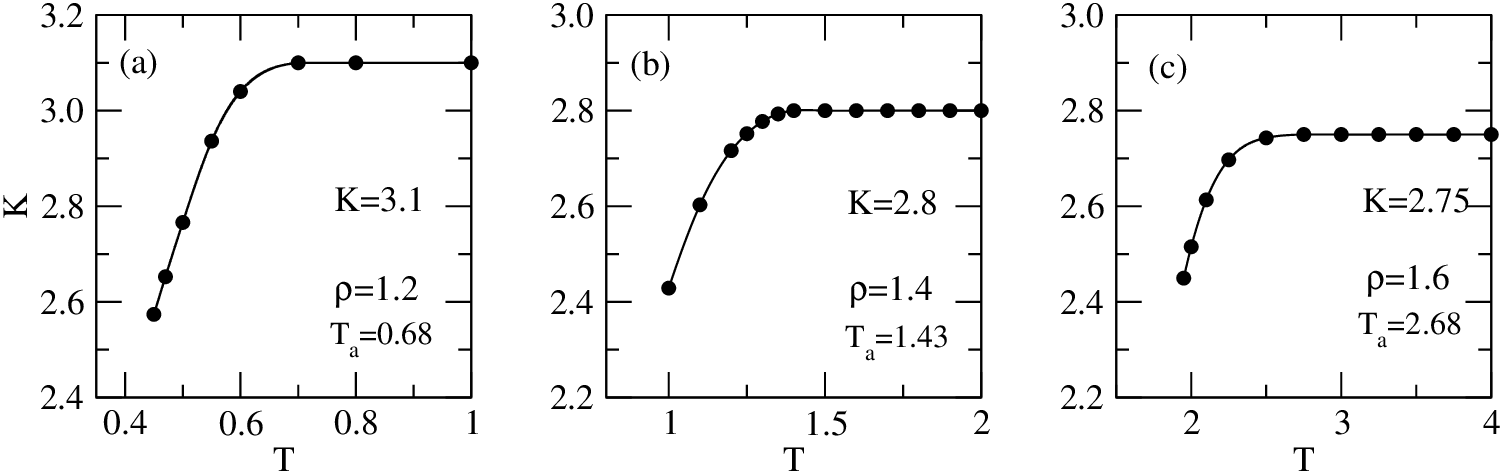}
\caption{Values of $K$ as a function of temperature $T$ found when $\psi=1$ was taken in 
calculating values of number of $s$-particles. The deviation of value of $K$ from its 
constant value at low temperature is due to the fact that taking $\psi=1$ at $T<T_{a}$
is not valid. Symbols represent calculated values and curves are least-square fit.}
\vspace{2cm}
\label{fig-2}
\end{figure*}

\begin{figure*}[]
\vspace{2cm}
\includegraphics[scale=0.5]{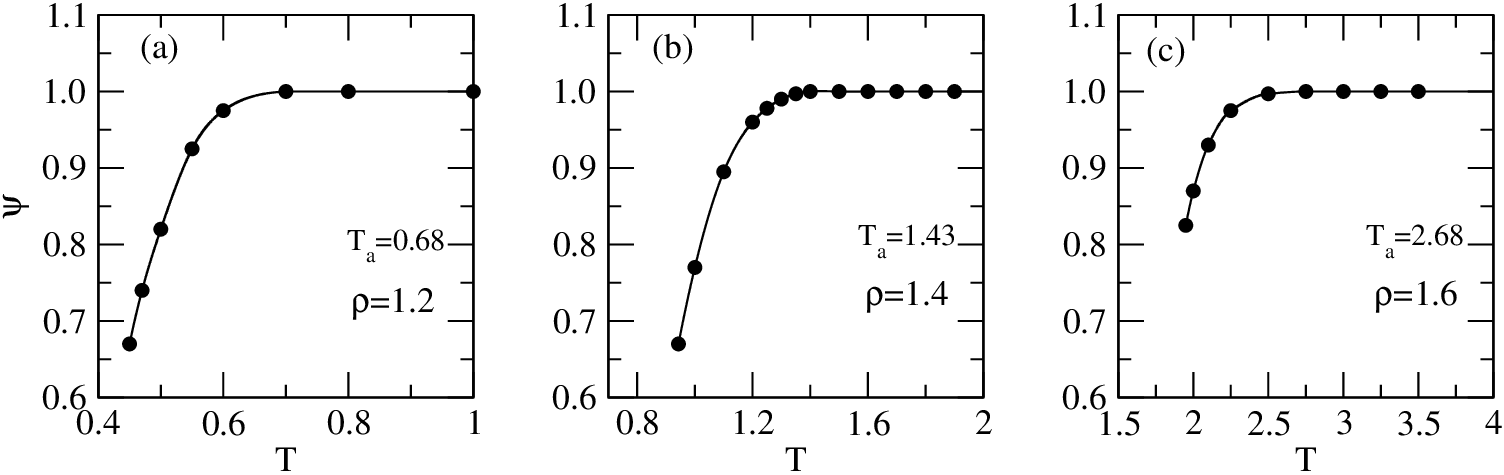}
\caption{Values of $\psi(T)$ as a function of temperature $T$ at different densities.
Symbols represent calculated values and curves are least-square fit.}
\vspace{-0.5cm}
\label{fig-3}
\end{figure*}

In order to determine value of $K$
we first take $\psi=1$ in Eqs. (\ref{grs}-\ref{nst}) and calculate $n^{(s)}(T)$ at
different temperatures. Values of $n^{(s)}(T)$ are then used in Eq. (\ref{Sc}) to
calculate $K$. In Fig. $2$ we plot $K$ vs $T$ at densities $\rho=1.2, 1.4$ and $1.6$.
In all the cases we find that $K$ is constant above a temperature  denoted as $T_{a}$;
both  $K$ and $T_{a}$ depend on $\rho$. However, for $T<T_{a}$, $K$ deviates from its constant value;
this we attribute to the fact that taking $\psi=1$ for $T<T_{a}$ is not valid.
As argued above, the value of $\psi$ is expected to decrease from its high{\color{red}-}temperature value 
on cooling below $T_{a}$ due to the increasing role of energy driven processes which counter the 
escape of particles from shells. We have more to say about the constant $K$ in Sec.~\ref{Discussion}, 
where we calculate its value from the data of $\tau_{\alpha}$ 
and show that it is indeed temperature independent and equal to the high temperature
value shown in Fig.~\ref{fig-2}.

\begin{center}
\begin{scriptsize}
\begin{table}[b]
\caption{Values of constant $K$, $T_{a}$, $ T_{onset} $ and
$ T_{mc} $ at different densities.}

\label{LJ-Data}
\begin{ruledtabular}
\begin{tabular}{>{\scriptsize}c>{\scriptsize}c>{\scriptsize}c>{\scriptsize}c>{\scriptsize}c>{\scriptsize}c>{\scriptsize}c>{\scriptsize}c>{\scriptsize}c>{\scriptsize}c>{\scriptsize}c}
\small

$ \rho $ & & $ K $ &   $ T_{a} $  & $ T_{onset} $  & $ T_{mc} $  \\
\hline

 $  1.2 $  &  &  $ 3.10 $  &  $ 0.68 $  &  $ 0.77 $  &  $ 0.43 $   \\
 $  1.4 $  &  &  $ 2.80 $  &  $ 1.43 $  &  $ 1.50 $  &  $ 0.93 $   \\
 $  1.6 $  &  &  $ 2.75 $  &  $ 2.68 $  &  $ 2.86 $  &  $ 1.76 $   \\
\end{tabular}
\end{ruledtabular}

\end{table}
\end{scriptsize}
\end{center}

Assuming that $K$ remains constant at all temperatures and has value
determined from high temperature result plotted in Fig. $2$ and listed in Table \ref{LJ-Data}, 
we determine $n^{(s)}(T)$ from Eq. (\ref{Sc}). The known values of $n^{(s)}(T)$ allow us 
to find temperature dependence of $\psi$ from Eqs. (\ref{grs}-\ref{nst}).
We plot $\psi(T)$ vs $T$ in Fig. $3$. We note that in all cases $\psi=1$ for $T>T_{a}$
but decreases rather sharply for $T$ less than $T_{a}$. Values of $T_{a}$ for the three 
densities are given in Table \ref{LJ-Data}. In this table we also list, for comparison sake, values
of ``onset temperature'' $T_{onset}$ and mode-coupling temperatures $T_{mc}$.
The $T_{onset}$ is defined as the crossing temperature of pair and excess entropies \cite{Banerjee}
and $T_{mc}$ is found by fitting data of $\tau_{\alpha}$ in a
power law form, $\tau_{\alpha}\propto(T-T_{mc})^{-\gamma}$ predicted by mode-coupling
theory (MCT) \cite{Gotze}. According to MCT, $T_{mc}$ is a temperature at which $\tau_{\alpha}$,
diverges; a prediction which is not observed. The temperature $T_{a}$ falls inbetween 
$T_{onset}$ and $T_{mc}$ and seems to separate a high temperature region where 
slowing down of dynamics is slower from a low temperature region where slowing down
of dynamics is faster. 

\begin{figure}[t]
\includegraphics[scale=0.5]{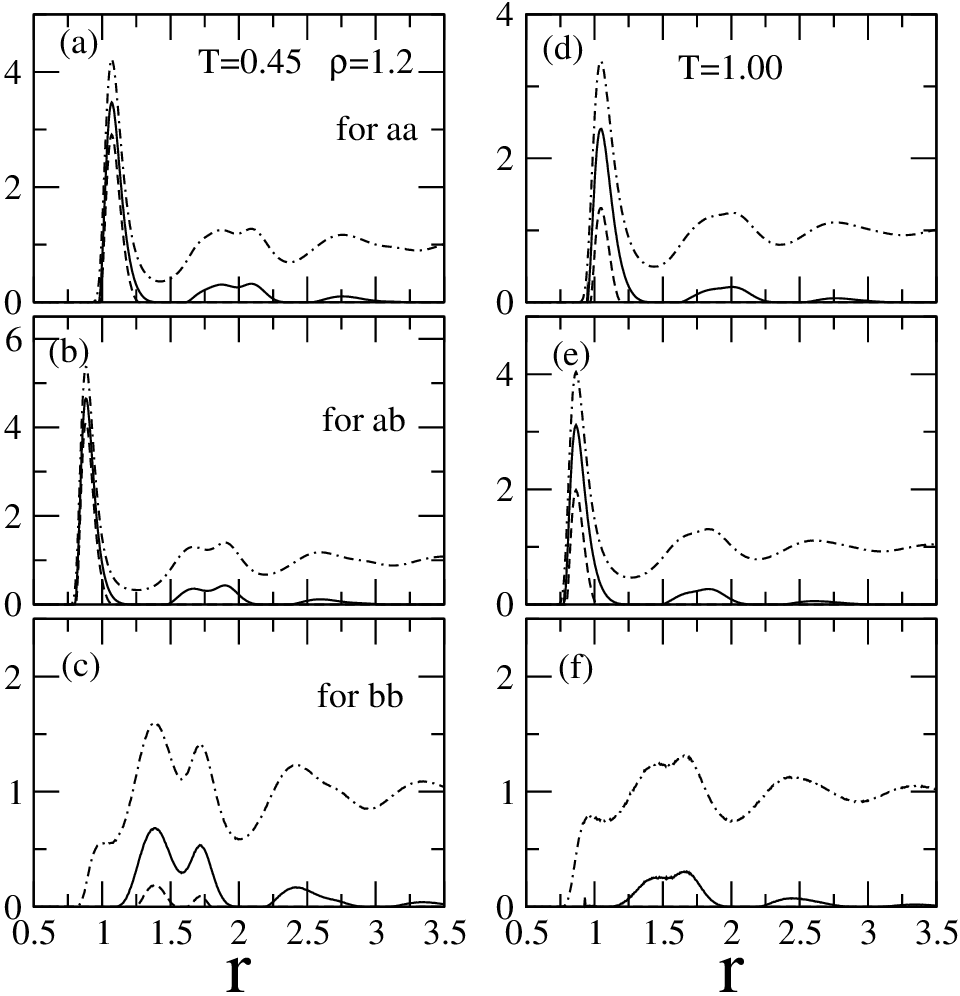}
\caption{Comparision of values of $g_{\alpha\gamma}(r)$ (dot-dashed line), $g_{\alpha\gamma}^{(b)}(r)$ (full line)
and $g_{\alpha\gamma}^{(s)}(r)$ (dashed line) at two temperatures
$T=0.45$, $1.0$ and the density $\rho=1.2$.}
\label{fig-4}
\end{figure}

\begin{figure}[t]
\includegraphics[scale=0.3]{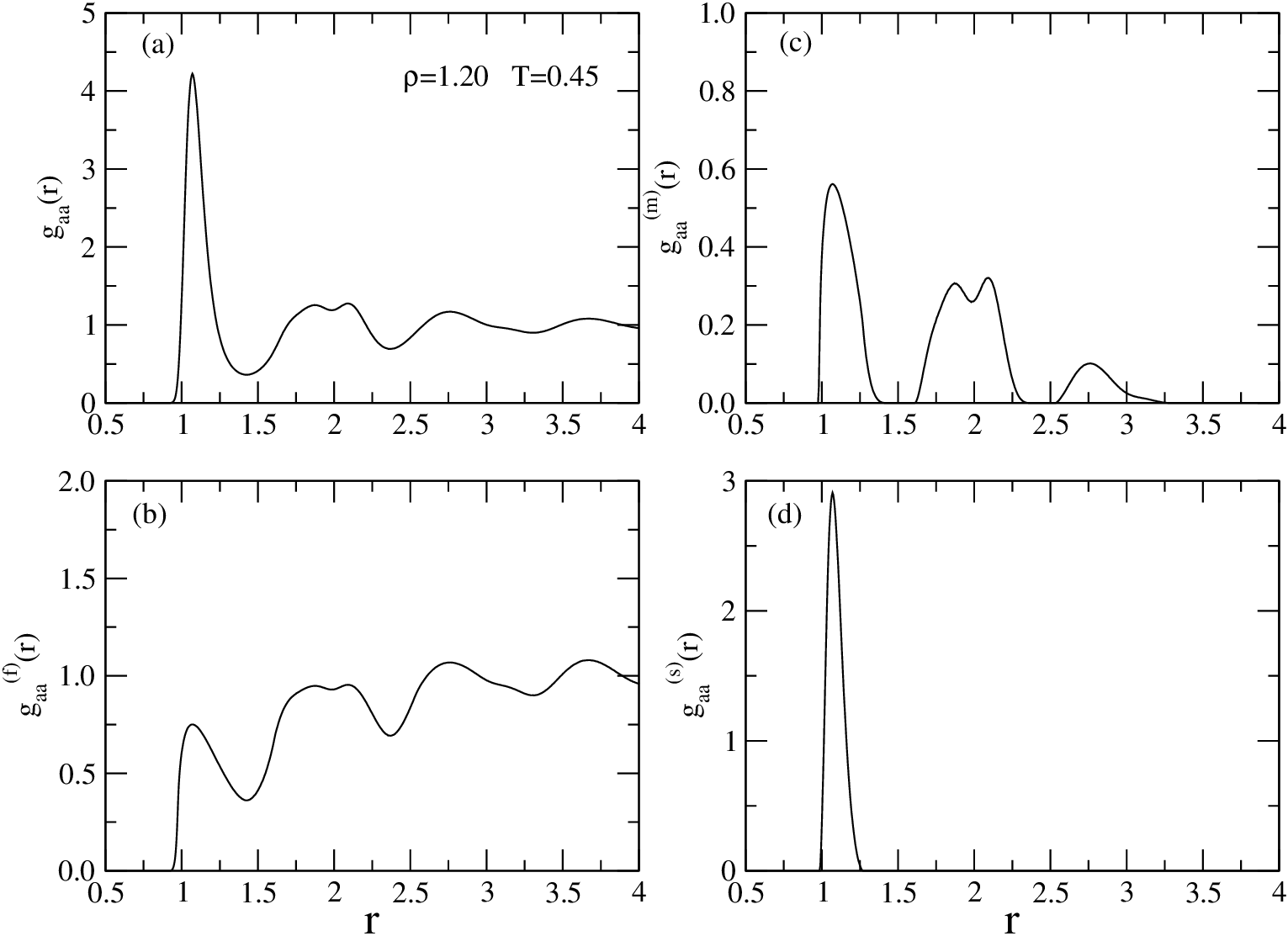}
\caption{Plot of $g_{aa}(r)$, $g_{aa}^{(f)}(r)
=g_{aa}(r)-g_{aa}^{(b)}(r)$, $g_{aa}^{(m)}(r)=g_{aa}^{(b)}(r)-g_{aa}^{(s)}(r)$ and 
$g_{aa}^{(s})(r)$ as a function of $r$ at $\rho=1.2$ and $T=0.45$.}
\label{fig-5}
\end{figure}

\begin{figure}[]
\includegraphics[scale=0.34]{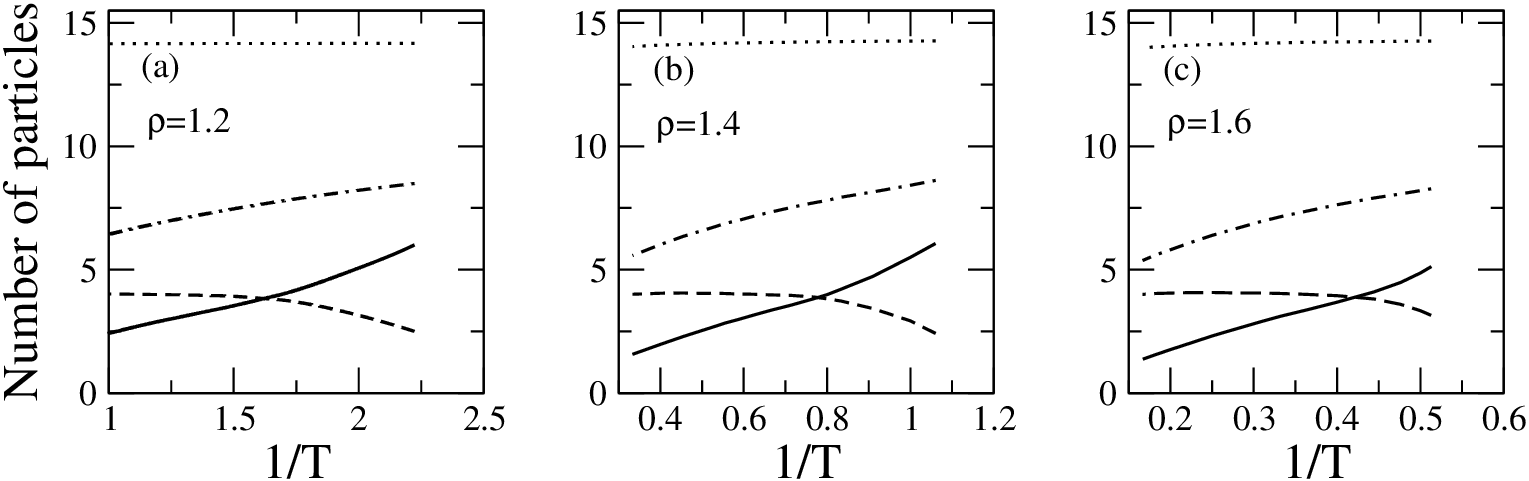}
\caption{Value of number of total $(n_{1}^{(t)})$ (dotted line), bonded 
$(n_{1}^{(b)})$ (dot-dashed line), metastable $(n_{1}^{(m)})$ (dashed line) and stable 
$(n_{1}^{(s)})$ (full line) particles occupying the first coordination 
shell as a function of $(1/T)$.}
\label{fig-6}
\end{figure}

We now use value of $\psi$ found at a given $T$ and $\rho$ in equations derived above 
to calculate different parts of $g_{\alpha\gamma}(r)$ and number of $m$- and 
$s$-particles. In Fig. $4$ we plot $g_{\alpha\gamma}(r)$, $g_{\alpha\gamma}^{(b)}(r)$
and $g_{\alpha\gamma}^{(s)}(r)$ as a function of distance $r$ for two temperatures
$T=0.45$ and $1.0$ and the density $\rho=1.2$ to show their temperature dependence.
In Fig. $5$ we explicitly show the  spatial range of $g_{aa}(r)$, $g_{aa}^{(f)}(r)
=g_{aa}(r)-g_{aa}^{(b)}(r)$, $g_{aa}^{(m)}(r)=g_{aa}^{(b)}(r)-g_{aa}^{(s)}(r)$ and 
$g_{aa}^{(s)}(r)$ for species $a$ at $\rho=1.2$ and $T=0.45$. From the figure one notes that
while $g_{aa}^{(s)}(r)$ is confined in the first shell with length scale of the order
of one particle diameter, $g_{aa}^{(m)}(r)$ extends to several shells with
length scale of the order of several particle diameter. In this context it is 
important to realize the role of $\psi$; as $\psi$ decreases on cooling the liquid,
contributions to $g^{(s)}(r)$ starts coming from other shells extending the associated
length scale. At a temperature where $\psi$ becomes zero all bonded particles become 
$s$-particles and length scale of $g^{(s)}(r)$ will be same as that of $g(r)-1$.

In Fig. $6$ we plot number of total particles $(n_{1}^{(t)}(T))$, bonded particles
$(n_{1}^{(b)}(T))$, $m$-particles $(n_{1}^{(m)}(T))$ and $s$-particles 
$(n_{1}^{(s)}(T))$ occupying the first  shell as a function of inverse of
the temperature $(1/T)$. We note that at high temperatures most particles are free,
while few are $m$-particles and very few are $s$-particles.
As the system is cooled. $n_{1}^{(m)}$ remains almost constant
but $n_{1}^{(s)}$ increases though slowly upto $T=T_{a}$. But for $T<T_{a}$,
$n_{1}^{(m)}(T)$ decreases while $n_{1}^{(s)}(T)$ increases with increasing
rate at the cost of both free and $m$-particles.
As stated above this rate will rapidly increase on further lowering of temperature
resulting in a rapid increase in the number of $s$-particles. To have a precise nature of this 
increase we need to have data  of $g_{\alpha\gamma}(r)$ at lower temperatures.

\section{Calculation of the potential energy barrier and the relaxation time}

The potential energy barrier (activation energy) to relaxation, as stated above,
is equal to the energy with which a particle is bonded with $s$-particles.
Thus
\begin{subequations}
\renewcommand{\theequation}{\theparentequation.\arabic{equation}}
\begin{eqnarray}\nonumber
\beta E^{(s)}(T,\rho) &=& 4\pi \sum_{i}\sum_{\gamma}x_{\gamma}\rho_{\gamma} \int_{r''_{il}}^{r''_{ih}} 
[\beta w_{\alpha\gamma}^{(iu)}-\psi(T)- \beta w_{\alpha\gamma}^{(i)}(r)] \\
&& \ \times g_{\alpha\gamma}^{(is)}(r) r^2 \mathrm{d}r ,
\label{energy}
\end{eqnarray}
where energy is measured from the effective barrier 
$\beta w_{\alpha\gamma}^{(iu)}-\psi(T)$. In Fig. $7$ we plot values of $\beta E^{(s)}$
vs $1/T$  for different densities. In all the cases we see sharp rise in $\beta E^{(s)}$
below $T_{a}$.

The energy $\beta E^{(s)}$ can be considered as the activation energy in the 
Arrhenius law,
\begin{equation}
\tau_{\alpha}(T,\rho) = \tau_{0} \exp{[\beta E^{(s)} (T,\rho)]}.
\label{tau_E}
\end{equation}
\end{subequations}
In  Fig. $7$ we compare calculated results with  values found from computer
simulations \cite{ABanerjee, Tarjus} for $\rho=1.2$, $1.4$ and $1.6$.
In all the cases we find very good agreement between calculated and 
simulation values.

\begin{figure}[t]
\includegraphics[scale=0.4]{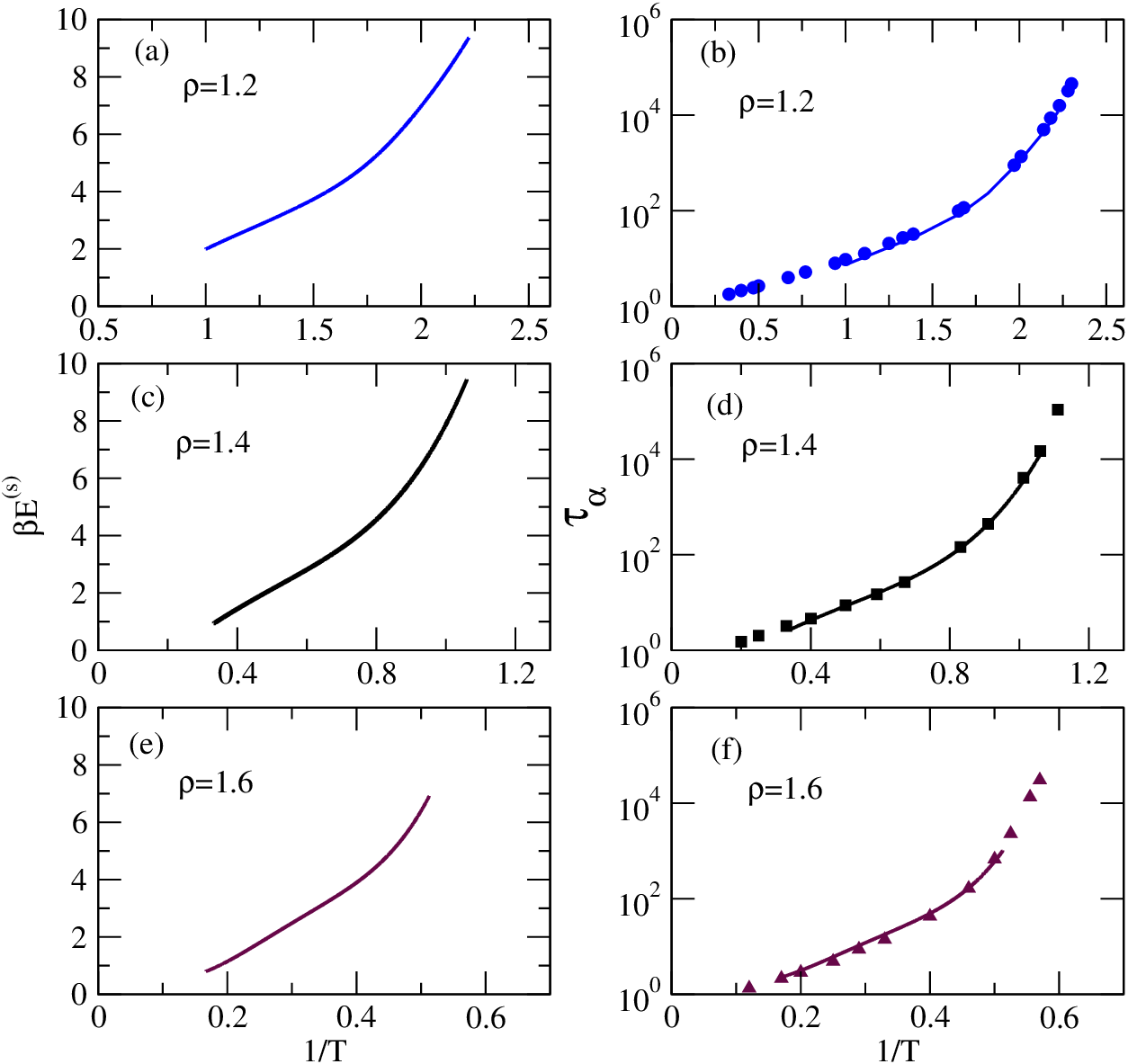}
\caption{Values of activation energy $\beta E^{(s)}$ for the relaxation and 
the relaxation time as a function of $1/T$ at three densities. Symbols
denote simulation values and curves denote calculated values.}
\label{fig-7}
\end{figure}

\section{The temperature $T_{a}$ and the density scaling}

\begin{figure}[t]
\includegraphics[scale=0.3]{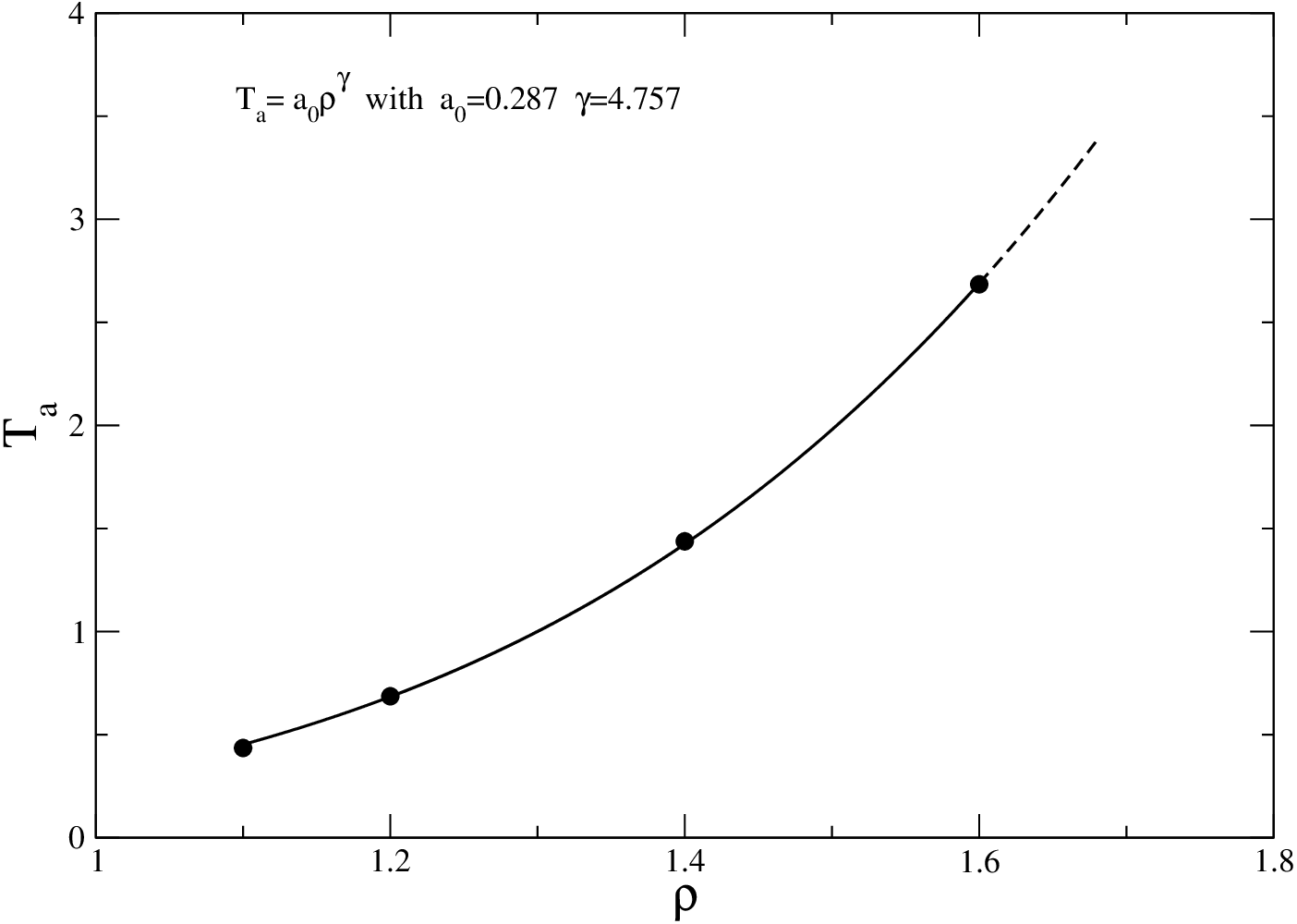}
\caption{Dependence of the temperature $T_{a}$ on density $\rho$ is shown. Full circles are calculated
values and curve represents the fit $T_{a}=a_{0}\rho ^{\gamma}$ with value of exponent $\gamma=4.757$.
The value of $T_{a}=0.435$ at $\rho=1.1$ was found from plotting $\tau_{\alpha}$ values on the 
master curve.}
\label{fig-8}
\end{figure}

\begin{figure}[t]
\includegraphics[scale=0.33]{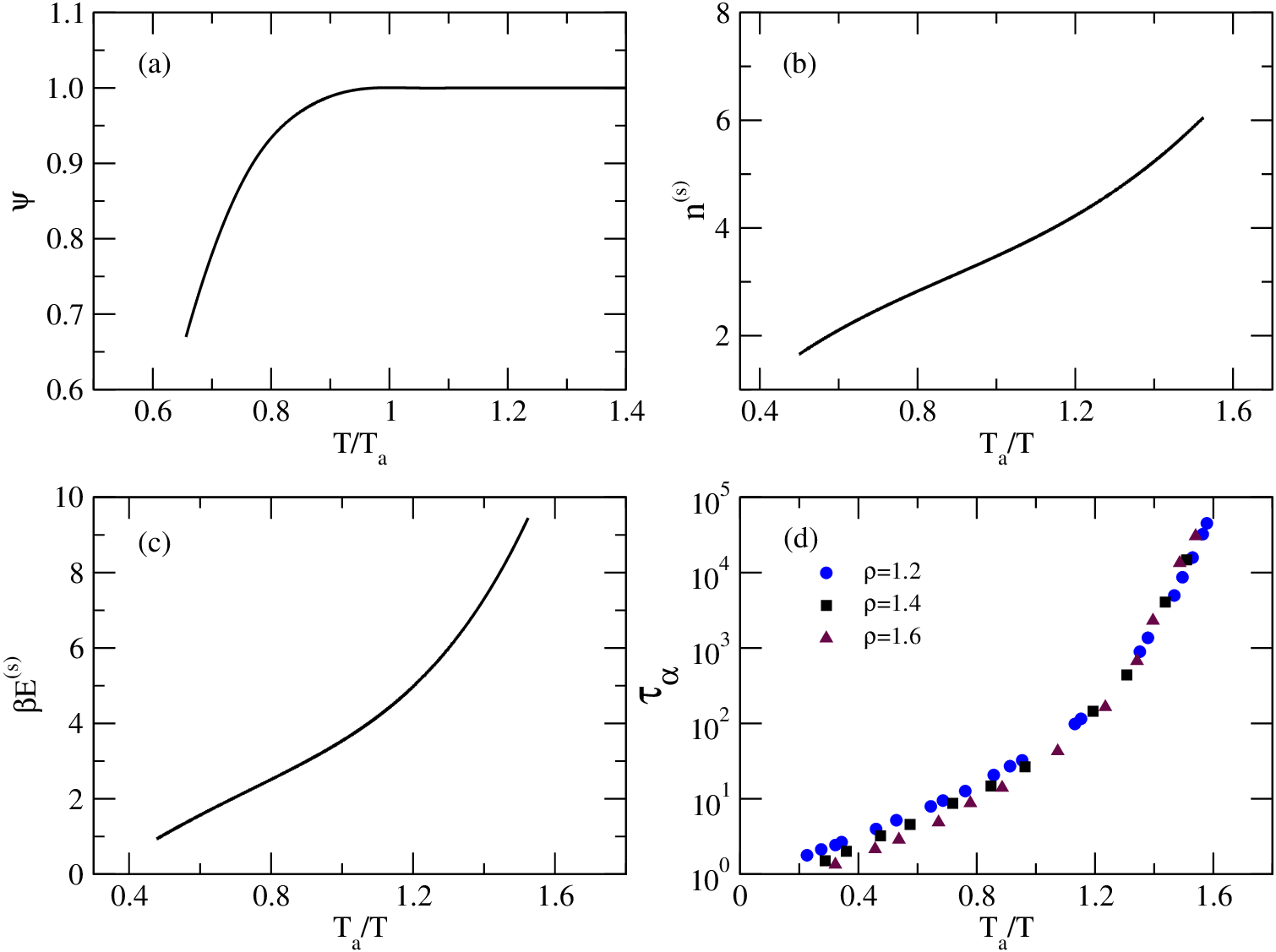}
\caption{Collapse of the data of $\psi$, $n^{(s)}$, $\beta E^{(s)}$ and $\tau_{\alpha}$ at densities
$\rho=1.2$ to $\rho=1.6$. In $(a)$ $\psi$ is plotted as a function of $T/T_{a}$ whereas in $(b)$,
$(c)$ and $(d)$ values of $n^{(s)}$, $\beta E^{(s)}$ and $\tau_{\alpha}$ are, respectively plotted 
as a function of $T_{a}/T$. This collapse of data on master curves shows that $T_{a}$ is a characteristic 
temperature of supercooled liquids and intimately connected with fluctuations embedded in the system.}
\label{fig-9}
\end{figure}

The density dependence of $T_{a}$ is shown in Fig. \ref{fig-8}. In the 
figure full circles denote  calculated values and the curve represents a
fit with a power law form; $T_{a}=a_{0} \rho ^{\gamma}$ with $a_{0}=0.287$
and $\gamma=4.757$. If we renormalize the temperature $T$  
by $T_{a}$ and plot values of $\psi(T)$
as a function of $T/T_{a}$ we find a very good collapse data of $\psi$ at
different densities as shown in Fig. \ref{fig-9}$(a)$. Similarly when we plot 
$n^{(s)}(\rho,T)$, $\beta E^{(s)}(\rho,T)$ and 
$\tau_{\alpha}(\rho,T)/\tau_{0}$ as a function of $T_{a}/T$ we find a very 
good collapse on master-curves as shown in Fig. \ref{fig-9}$(b)$, $(c)$ and $(d)$, respectively.

Scaling of thermodynamics and dynamic properties in terms of $\rho^{\gamma}/T$
was argued to be a property of ``strongly correlating liquids'' \cite{Dyre}. These liquids have 
strong correlations between their constant-volume equilibrium fluctuations of potential
energy $U(t)$ and virial $W(t)=-{1/3}\sum_{i} \vec{r_{i}}\cdot{\Delta_{\vec{r_{i}}}} U(\vec{r_{1}},\cdots,\vec{r_{N}})$
where $U(t)$ is the total potential energy at time $t$ and 
$\vec{r_{i}}$ is the  position of particle $i$ at time $t$.
The correlation is characterized by a single parameter $\Gamma$,
defined by a linear fit through a scatter plot of time fluctuations
of $U$ and $W$. For the strongly correlating liquids 
$\Gamma\simeq\gamma$. The parameter $\gamma$ found from the slope of
correlation plot as shown in Ref.~\cite{Tarjus} has some arbitrariness;
it varies between $4.5$ and $5.1$ for the model of Eq. (\ref{potential}).
If we ignore this variation, the  value can be considered to be 
in good agreement with the one found from the density dependence
of $T_{a}$. This suggests that $T_{a}$ is related to the ``hidden scale
invariance'' \cite{Dyre, Schroder} and is a result of strong $WU$ correlations.

\section{Discussions\label{Discussion}}

\begin{figure}[t]
\includegraphics[scale=0.3]{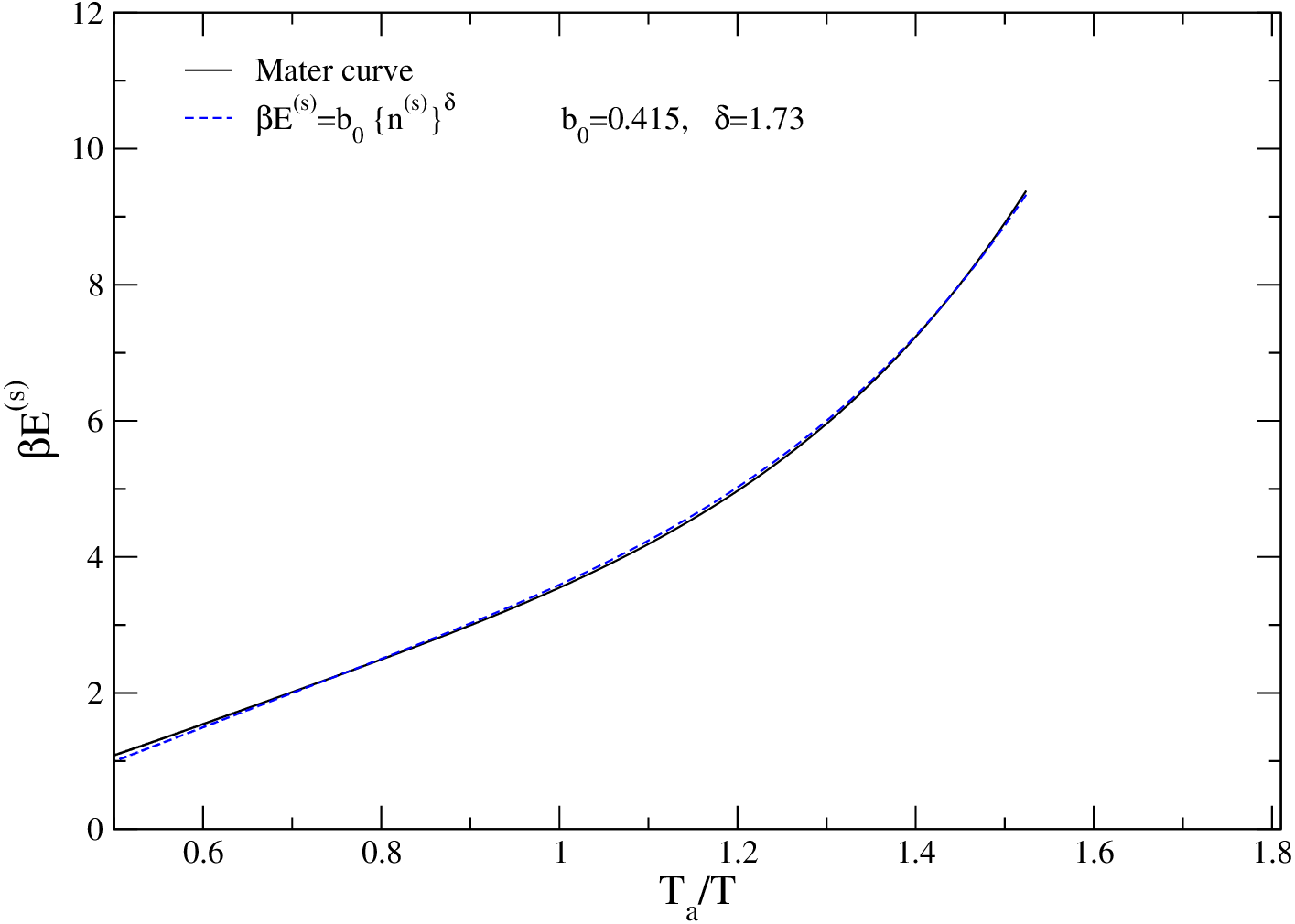}
\caption{Plot of the reduced activation energy $\beta E^{(s)}$ of relaxation as a function of
$T_{a}/T$ at densities $\rho=1.2, 1.4$ and $1.6$. The full line represents values found from 
the calculation and the dashed line represents a fit $\beta E^{(s)}=b_{0}\lbrace n^{(s)}\rbrace^{\delta}$
where $b_{0}=0.415$ and $\delta=1.73$ and  $n^{(s)}$ is the number of bonds in a CRC (see Fig. \ref{fig-9}(b)).}
\label{fittingE}
\end{figure}

\begin{figure}[t]
\includegraphics[scale=0.33]{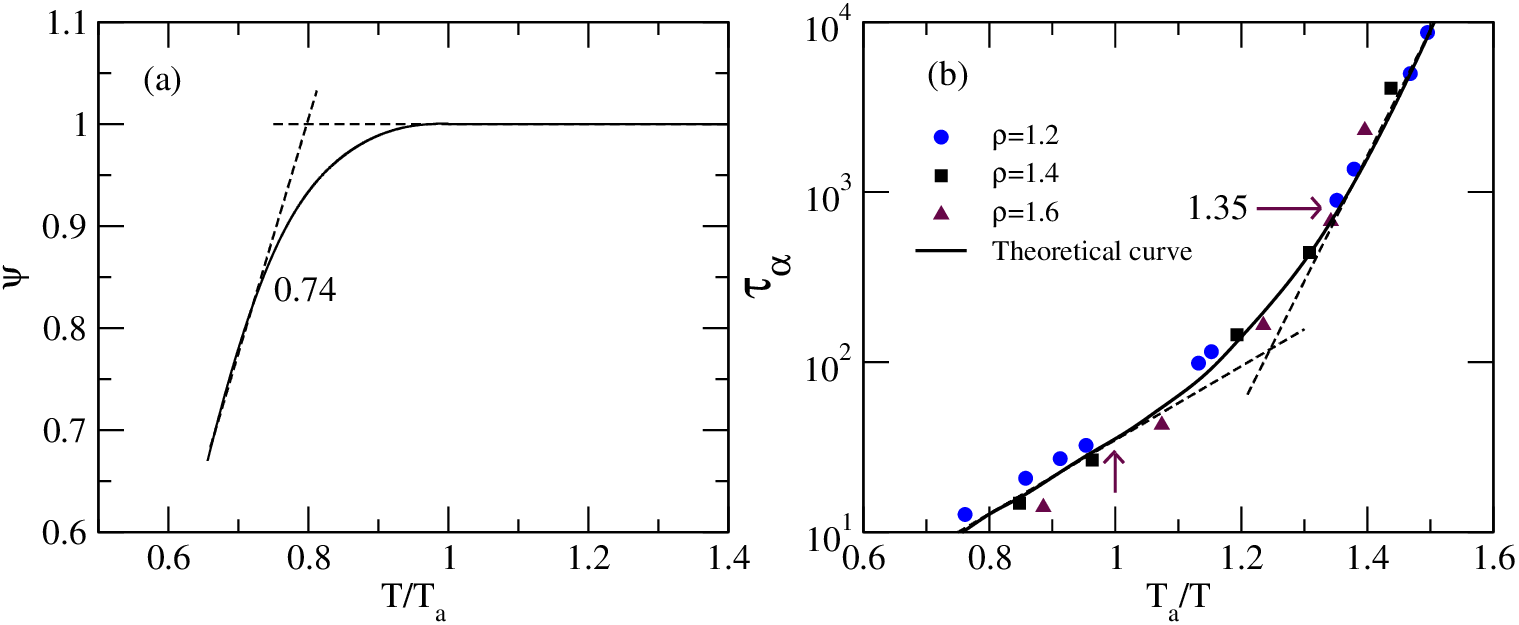}
\caption{Figures show the crossover region of $\psi (T)$ (in (a)) and $\tau_{\alpha}$ (in (b)) 
from the high temperature behavior to the low temperature behavior on cooling.
In both cases the crossover region is in the range $0.74\lesssim T/T_{a} \lesssim 1$.
The straight lines (dashed line) are drawn to separate out the crossover region.}
\label{fittingcrossover}
\end{figure}

\begin{figure}[t]
\includegraphics[scale=0.3]{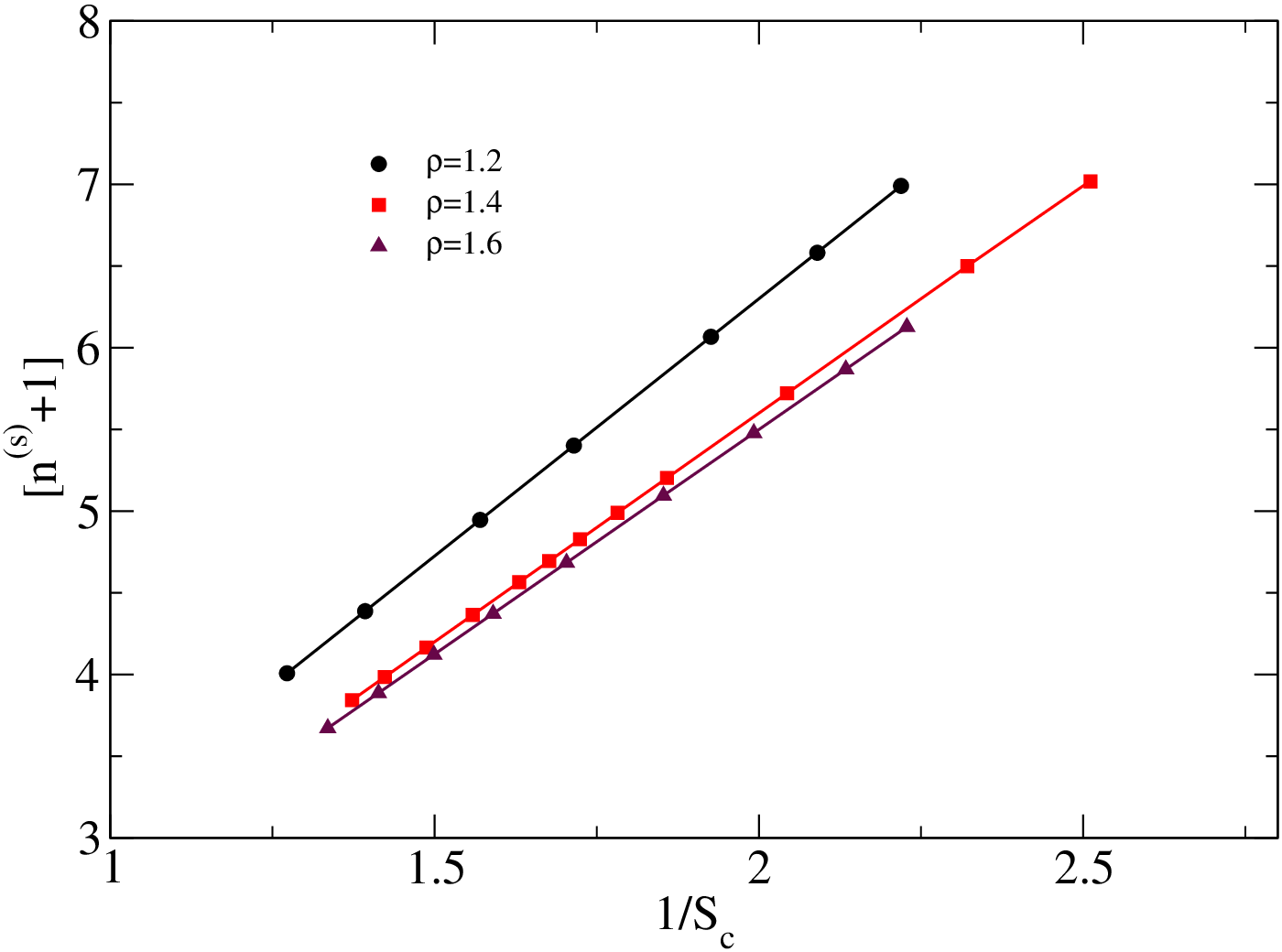}
\caption{Plot of number of particles $n^{(s)}+1$ in a CRC vs $1/S_{c}$ at densities $\rho=1.2, 1.4$ and $1.6$.
Lines show that the number $n^{(s)}+1$ is inversely proportional to the configurational
entropy $S_{c}$ and the proportionality constant is temperature independent. Values of constants determined from the slope of these curves are same as the one given in Table \ref{LJ-Data}}
\label{particlevsentropy}
\end{figure}

The theory described in this paper brings out several underlying features of the dynamics of fragile
supercooled liquids. The local structural order conceptualized as something which has to be a ``very
subtle and hidden to the pair correlation function'' is shown to be the cooperatively reorganizing
cluster (CRC) formed by a central particle with its neighbors of localized ($s$-) particles and is 
determined from the radial distribution function, $g(r)$ by including momentum distribution in its
definition. The CRC is, however, not the same as the
Adam-Gibbs, CRR (cooperatively rearranging region) which is taken to be a compact structure characterized
by a static length \cite{Adam}. But, in the CRC, particles are distributed in coordination shells 
surrounding the central particle (see Fig.~\ref{fig-1}) and share the space with other (mobile) particles.
As suggested earlier the CRC structure can be porous; 
the number of particles in it can increase without increasing its size.

The cooperativity of relaxation is defined in terms of the number of particles (or bonds) $n^{(s)}$
which are connected with the central particle in a CRC. For an event of  relaxation to take place
these bonds have to rearrange irreversibly; the energy involved in this process is the effective
activation energy $\beta E^{(s)}$ of relaxation. As the system is cooled, both the number $n^{(s)}$
and the energy of each bond increase; the combined effect makes the $\beta E^{(s)}$ to increase rapidly
as temperature is lowered. A fit of (collapsed) data of $\beta E^{(s)}$ with a power law form
$\beta E^{(s)}=b_{0}\lbrace n^{(s)}\rbrace^{\delta}$ with $b_{0}=0.415$
and $\delta=1.73$ is shown in Fig. \ref{fittingE}.

The parameter $\psi(T)$ which is introduced to measure the effect that the bath creates to stabilize
the size of CRC gives insight into the processes underlying the slowing down of dynamics. The value of 
$\psi(T)$ takes a turn from its high temperature value of $1$ at $T=T_{a}$ and starts decreasing as $T$
is lowered. There is a crossover region (see Figs. \ref{fig-3} and \ref{fig-9}) which separates the high temperature behavior
from the low temperature behavior. Exactly same behavior is seen (see Figs. \ref{fig-7} and \ref{fig-9})
in $\tau_{\alpha}$; a crossover from the high temperature $(T>T_{a})$ behavior to the low temperature
$(T<T_{a})$ behavior takes place in the same way as it happens in the case of $\psi$.
This is more clearly seen in Fig. \ref{fittingcrossover} where we marked the two regions with straight lines
to separate the crossover region. This brings forth the underlying cause for the temperature
dependence of $\tau_{\alpha}$. The crossover region marks the change from the dominance of the entropy
driven process to the dominance of the energy driven processes in the system on cooling.

Both $\psi$ and $T_{a}$ depend on details of the inter-particle interactions. Value of temperature
$T_{a}$ is sensitive to the attraction in the potential; its value would increase on increasing the
attraction. Since the rate of slowing down of dynamics increases for $T<T_{a}$ the two systems with the same
repulsion but a different attraction in the pair potential would show very different dynamics while
$g(r)$ of the two systems may appear similar. This will be investigated in detail in our next 
publication.

We wish to emphasize that though Eq.~(\ref{Sc}) lacks theoretical 
rigour \cite{Bouchaud, ACavagna} it, as has been shown in Sec.~\ref{Theory}$B$, correctly describes the 
relationship between number of particles in a CRC
and the configurational entropy. Here, we adopt a method different from the one described in Sec.~\ref{Theory}$B$ 
and show that the number of particles in a CRC are inversely proportional to the
configurational entropy and calculate the constant $K$. In particular, we now use data of 
$\tau_{\alpha}/\tau_{0}$ determined from simulations in Eq.~(\ref{tau_E}) to calculate 
$\beta E^{(s)}$ at different temperatures and then use these values in Eqs.~(\ref{grs}-\ref{nst}) to calculate values of 
$\psi (T)$ and $n^{(s)}(T)$ \cite{Ankit}. The plot of $n^{(s)}(T)+1$ vs $1/S_{c}(T)$ in Fig.~\ref{particlevsentropy} for the three
densities  show that $n^{(s)}(T)+1$ is indeed inversely proportional to $S_{c}$ and the proportionality
constant $K$ is temperature independent. Values of $K$ found from the slope of curves of different 
densities are same as the one given in Table~\ref{LJ-Data}.

In summary, by including momentum distribution in the definition of the radial distribution function $g(r)$
and using the information of the configurational entropy $S_{c}$, we calculated the number of free, metastable,
and stable particles  distributed in coordination shells surrounding a central particle at different temperatures
and densities. It is shown that in supercooled liquids slowing down of dynamics is due to the emergence of 
cooperatively reorganizing clusters (CRCs) in which the 
central particle forms (nonchemical) ``stable bonds'' with neighboring stable particles. The number of bonds
with which the central particle is bonded with neighbors defines the cooperativity of relaxation. For an event of
relaxation to take place these bonds have to reorganize irreversibly. The energy involved in this process is the 
effective activation energy $\beta E^{(s)}$ of relaxation. The number of bonds and the energy of each bond of the 
CRC increase in lowering the temperature and increasing the density. When $\beta E^{(s)}$ is substituted in the 
Arrhenius law, a super-Arrhenius feature emerges. Results found for $\tau_{\alpha}$ for the Kob-Anderson model system 
are in very good agreement with simulation results.  The temperature dependence of $\tau_{\alpha}$
is explained, in terms of the parameter 
$\psi(T)$ which measures the effect of fluctuations embedded in the bath on stabilizing the size of CRC.

\section*{Acknowledgment}
A. S. acknowledges research fellowship from the Council of Scientific and Industrial Research, New Delhi, India. S.M.B will like to thank SERB, India for funding.

\end{document}